\documentclass[reprint,amssymb,amsmath, aps, nofootinbib]{revtex4-2}
\usepackage{graphicx}
\usepackage{bm} 
\usepackage{pxfonts}
\usepackage{mathpazo}
\usepackage{color}

\definecolor{purple}{rgb}{0.5,0,0.9}

\begin{document}

\title{Searching for gravitational wave echoes from black hole binary events in the third observing run of LIGO, Virgo, and KAGRA collaborations}


\author{Nami Uchikata$^1$} 
\email{uchikata@icrr.u-tokyo.ac.jp}

\author{Tatsuya Narikawa$^1$}
\email{narikawa@icrr.u-tokyo.ac.jp}

\author{Hiroyuki Nakano$^2$}
\email{hinakano@law.ryukoku.ac.jp}

\author{Norichika Sago$^3$}
\email{sago@tap.scphys.kyoto-u.ac.jp}

\author{Hideyuki Tagoshi$^1$}
\email{tagoshi@icrr.u-tokyo.ac.jp}
\author{Takahiro Tanaka$^{3,4}$}
\email{t.tanaka@tap.scphys.kyoto-u.ac.jp}
\affiliation{
$^1$Institute for Cosmic Ray Research, The University of Tokyo, Chiba 277-8582, Japan\\
$^2$Faculty of Law, Ryukoku University, Kyoto 612-8577, Japan \\
$^3$Department of Physics, Kyoto University, Kyoto 606-8502, Japan \\
$^4$Center for Gravitational Physics, Yukawa Institute for Theoretical Physics, Kyoto University, Kyoto 606-8502, Japan}

\begin{abstract}
Gravitational wave echo signals have been proposed as evidence for the modification of the spacetime structure near the classical event horizon.
These signals are expected to occur after the mergers of compact binaries as a sequence of weak pulse-like signals.
Some studies have shown evidence of the echo signals from several binary black hole merger events. 
On the other hand, the other studies have shown the low significance of such signals from various events in the first, second and third observing runs (O1, O2 and O3).
Our previous study also shows the low significance of echo signals from events in O1 and O2, though, we observe that more than half of the events have p-value smaller than 0.1 when the simply modeled waveform is used for the analysis.
Since there are only nine events appropriate for this analysis in O1 and O2, it is necessary to analyze more events to evaluate the significance statistically. 
In this study, we search for echo signals from binary black hole events observed during O3 operated by LIGO, Virgo and KAGRA collaborations.
We perform the template-based search by using two different models for echo signal templates: simply modeled one and physically motivated one.
Our results show that the distributions of p-values for all events analyzed in this study are consistent with the noise distribution.
This means that no significant echo signals are found for both models from O3 events.
\end{abstract}

\date{\today}
\maketitle

\section{Introduction}

Successful gravitational wave observations from compact binary coalescences \cite{O1,catalog,gwtc2,gwtc2-1,gwtc3} have made gravitational waves a standard tool to investigate the spacetime structure.
Particularly, they have made it possible for us to directly test general relativity in the strong gravitational field \cite{ligo2,TGR-gwtc2,tgr-gwtc3}.
Although these tests have shown that the observational results are consistent with general relativity, properties of the remnant compact objects have been weakly constrained for the deviation from black holes.

Several studies have proposed compact object models as an alternative to black holes, such as a gravastar \cite{mazur} and the firewall \cite{almheiri}.
If the remnant objects of compact binary coalescences have the light ring and a surface instead of the event horizon, we might observe additional signals after the merger of compact binaries, which are called gravitational wave echoes \cite{cardoso,cardoso2,Cardoso:2017njb,cardoso4}.
These studies explain the following process.
Initially, we expect to observe the ringdown signals as a prompt emission after the merger, which is similar to the black hole's one.
Then, part of merger-ringdown signals will fall into the object, and iteratively reflected at the object's surface and the angular momentum barrier.
During the iterative reflections, the signals will gradually leak outside the potential barrier and be observed as late-time echoes, which will approach to the object's characteristic oscillations.
Similarly, some specific ``quantum black holes'' can be a candidate of the source of echo signals, since the event horizon can only absorb signals with specific discrete frequencies \cite{cardoso3,wang,oshita,Agullo:2020hxe}.
Therefore, the observation of echoes is strong evidence to claim that the near horizon structure is different from the classical one, predicted by general relativity.
However, to detect significant echo signals is challenging, since we do not know the true waveform of echoes even if they exist and the amplitude is expected to be much weaker than the merger amplitude.
For example, the amplitude of echoes is constrained to 15\% of the merger amplitude for GW150914 \cite{nielsen} and another study shows that necessary signal-to-noise ratio of the ringdown signal is $20 \sim 60$ to detect echoes \cite{Micchi:2020gqy}.

So far, evidence of echo signals from GW150914, GW151012, GW151226, and GW190521 has been reported by the model dependent search \cite{abedi,echoGW190521,Abedi:2022bph}.
On the other hand, other studies have shown the low significance of echo signals from various events \cite{Ashton, westerweck,nielsen,lo,TGR-gwtc2,uchikata,Wang:2020ayy}.
For the model independent search, no significant echo signals has been reported from events in the first, second and third observing runs (O1, O2, and O3) \cite{tsang,Tsang:2019zra,tgr-gwtc3,echoGW190521,Ren:2021xbe,Miani:2023mgl}, except for the binary neutron star merger event GW170817  \cite{Abedi:2018npz}.

For the model dependent search, we need to prepare a precise echo waveform model.
The simplest model for the echo signals is the repetition of the merger-ringdown waveform with a constant decay rate \cite{abedi}, which many studies listed above have used.
Later, several studies have updated the model by including physical effects \cite{mark,testa,maselli,nakano,Maggio:2019zyv,ytwang,Conklin:2021cbc}.
These models are constructed from the waves that are reflected at the surface. 
Some studies examine the behavior of the wave that falls into the object before it is reflected at the surface for more accurate modeling \cite{Sago:2020avw,Xin:2021zir,Ma:2022xmp}.

Although we have not found significant echo signals from O1 and O2 events previously \cite{uchikata}, our results have shown that five out of nine events have p-values less than $0.1$ when the simply modeled template proposed in Ref.~\cite{abedi} is used.
This non-uniform p-value distribution seemingly contradicts with the null hypothesis that the data do not contain echo signals.
However, since there were just nine events, to statistically judge the deviation from the null hypothesis, we must analyze more events.
To this end, in this study, we search for echo signals for black hole binary events observed during O3 operated by LIGO, Virgo, and KAGRA collaborations.
We analyze the data with two templates: simply modeled one \cite{abedi} and physically motivated one \cite{nakano}.
We use the former model to investigate the change of the non-uniformity of the p-value distribution when more events are analyzed.
We assume the latter model as a representative physically motivated model.
Then, we evaluate the significance of the possible echo signals with p-values and compare their distributions with the null hypothesis.

The paper is organized as follows.
In Section II, we explain the template waveforms used in our analysis.
In Section III, we describe the method of analysis to evaluate the significance of echo signals and the treatment of the data.
In Section IV, we show the significance of echo signals by the distributions of p-values of all events.
Conclusions and discussions are given in Section V. 
We present injection study of our analysis method in Appendix~\ref{injection},
and summarize our search region in Appendix~\ref{search}.

\section{Template waveforms}

In this study, we assume that the spacetime is entirely Kerr spacetime but a reflective surface is located at about Planck length away from the event horizon radius.
Because of the reflective surface, part of the merger-ringdown signals will be iteratively reflected between the surface and the angular momentum barrier of the Kerr spacetime.
Every time the signals reach at the potential barrier, part of them will transmit to infinity, therefore, we might observe series of weak pulse-like signals, i.e.~echoes.
Echo waveforms can be characterized by the reflection rates at the surface and at the potential barrier, and the time interval of echoes $\Delta t_{\rm echo}$, which corresponds to twice the proper distance between the surface and the potential barrier.
In this study, we assume that $\Delta t_{\rm echo} \lesssim O(1)$~sec, while recent study points out the possibility that $\Delta t_{\rm echo}$ can be billions of years for some models \cite{Zimmerman:2023hua}.

We assume a perfect reflection at the surface.
For the reflection rate at the potential barrier, we consider two cases: one with a constant reflection rate  (the Simple model) \cite{abedi}  and the other with a frequency dependent reflection rate by considering the effect of black hole perturbations (the BHP model) \cite{nakano}. 
We assume the BHP model as a representative model of physically motivated waveforms, while the Simple model gives smaller p-values than the BHP model in our previous study \cite{uchikata}.

We describe each template in the following subsections, where we use $c=G=1$ units with $c$ and $G$ being the speed of light and the gravitational constant, respectively.

\subsection{Simple model}
The Simple model assumes that echo signals consist of a repetition of a merger-ringdown waveform with an interval $ \Delta t_{\rm echo} $ and a constant decay rate $\gamma$, which is related to the refection rate at the potential barrier, expressed as 
\begin{equation}
\tilde{h}_{\rm simple} (f) =  \tilde{h}_0(f) \sum^{N_{\rm echo}}_{n=1} \gamma^{n-1} (-1)^{n-1} e^{- i \left (2\pi f \Delta t_{\rm echo} \right ) (n-1)}  \, .
\label{template1}
\end{equation}
Here, the tilde denotes the Fourier transform of the corresponding timeseries function and $N_{\rm echo}$ is the number of echoes.
The template assumes that the phase shift due to the reflection is fixed to $\pi$.
The interval $ \Delta t_{\rm echo} $ is determined by the remnant black hole spin $a$ and mass $M$ as 
\begin{equation}
\begin{split}
\Delta t_{\rm echo} = 2 \int_{r_++\Delta r} ^{r_{\rm max}} \frac{r^2+a^2}{r^2-2M r + a^2} dr \, ,
\end{split}
\end{equation}
where $r_{\rm max}$ is the peak of the angular momentum barrier and $\Delta r$ is the coordinate distance of the surface from the horizon radius $r_+$.
The main waveform $h_0(t)$ in Eq.~\eqref{template1} is given by cutting off the inspiral part of the best fit inspiral-merger-ringdown (IMR) waveform $h_{ \mbox{{\tiny IMR}} }^{\mbox {\tiny B}}(t)$ for each event as
\begin{equation}
\begin{split}
 h_0(t) &= \frac{1}{2}\left \{ 1 + \tanh \left [ \frac{1}{2}\omega (t-t_{\rm merger} -t_0) \right ] \right \} \\
     & \quad  \times h_{ \mbox{{\tiny IMR}} }^{\mbox {\tiny B}}(t), 
 \end{split}
 \label{seed}
\end{equation}
where $\omega$ is a representative frequency estimated around the merger time, $t_{\rm merger}$ is the merger time of a binary, and $t_0$ is a cutoff parameter.
Since $\omega$ and $t_0$ are rather insensitive to the analysis, we set $\omega$ as a constant value and $t_0 = -0.1\Delta t_{\rm echo}$.
We further fix the initial phase as 0.
The merger time can be determined by analyzing with $h_{ \mbox{{\tiny IMR}} }^{\mbox {\tiny B}}(t)$.
Therefore, for this template, we assume $( \Delta t_{\rm echo} ,\gamma)$ as search parameters.
The search region of $ \Delta t_{\rm echo}$ is calculated from the 90\% credible region of the remnant mass and spin of the binary black holes from parameter estimation.
\subsection{BHP model}
For the second model, we include all physical effects as much as possible.
The model uses a frequency-dependent reflection rate at the potential barrier, obtained by solving black hole perturbations \cite{nakano}, 
\begin{widetext}
\begin{equation}
\tilde{h}_{\rm BHP} (f) = \sqrt{1-R^2(f)}\, \tilde{h}_0(f)
\sum^{N_{\rm echo}}_{n=1} R(f)^{n-1} e^{- i \left (2\pi f \Delta t_{\rm echo}+ \phi \right ) (n-1)}  \, ,
\label{template2}
\end{equation}
\end{widetext}
where $\phi $ is the overall phase shift due to the reflections at the surface and the potential barrier.
Since the main frequency range of the model is concentrated around the black hole quasinormal mode frequency, where the frequency dependence of $\phi$ is weak, and  the linear term of the phase can be absorbed in  $\Delta t_{\rm echo}$, we approximate $\phi$  as a constant.

For the reflection rate, we use a fit for $0.6 \le \chi \le 0.8$ from numerical calculations, where $\chi =a/M$.
The fit is given as
\begin{widetext}
\begin{equation}
R(f) \approx
 \frac{1+e^{-300(2 \pi M f+0.27-\chi) } +  e^{-28(2  \pi M f- 0.125- 0.6 \chi) } }{ 1+e^{-300(2  \pi M f+0.27-\chi) } +  e^{-28(2  \pi M f- 0.125- 0.6\chi) } + e^{19(2  \pi M f- 0.3- 0.35\chi) } } .
\label{Rf}
\end{equation}
\end{widetext}
Unlike the Simple model, we also optimize the initial phase by constructing two orthogonal templates, which are used in the quasinormal mode analysis of black holes \cite{Nakano:2003ma,Nakano:2004ib,Uchikata:2020wsp}.

For this template, we assume $(a,M,\phi)$ as search parameters. 
Since $\Delta t_{\rm echo}$ is determined by $(a,M)$, $\Delta t_{\rm echo}$ is also varied during the search.
The search regions of $ (a,M)$ are given from the 90\% credible region of the remnant mass and spin of the binary black holes from parameter estimation.

\begin{figure}[ht]
\includegraphics[scale=0.55]{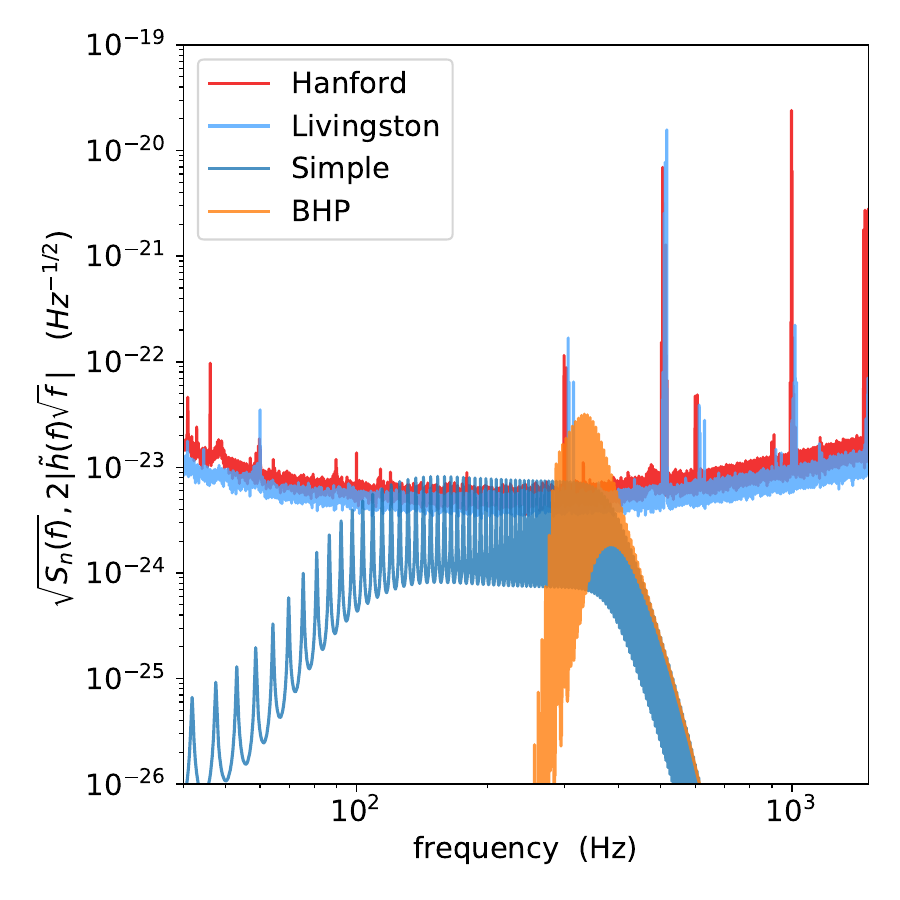}
\caption{Spectra of the best-fit echo templates of the Simple (blue) and BHP (orange) models for GW190412 are shown.
Detector's noise for this event for Hanford (red) and Livingston (cyan) are also shown.  
Best fit values are $(\Delta t_{\rm echo}, \gamma) = (0.1129\,s , 0.9)$ for the Simple model and $(a,M,\phi) = (0.68, 45.2 \,M_{\odot}, 2.356)$ for the BHP model.}
\label{template}
\end{figure}

In both templates, the best fit waveform $ h_{ \mbox{{\tiny IMR}} }^{\mbox {\tiny B}}(t)$ is given by SEOBNRv4 \cite{Bohe:2016gbl} or NRsur7dq4 \cite{Varma:2019csw} with mass and spin parameters chosen from the maximum likelihood of the results of parameter estimation.
Figure \ref{template} shows an example of the spectra of best-fit templates and detector noise for GW190412.
Because of the frequency dependent reflection rate, the lower frequency region of the BHP model is strongly suppressed compared with the Simple model.
We also consider the dominant mode only, since the amplitude of higher multipole modes are expected to be much smaller than the dominant mode.

\section{Method of analysis}
We first search for possible echo signals around the theoretically predicted time after the black hole binary mergers.
Then, we evaluate the significance of the echo signal as p-values in comparison with the background data.
\subsection{Search for echo signals}
We use matched filtering to search for echo signals.
The signal-to-noise ratio (SNR) $\rho$ for matched filtering is defined by 
\begin{equation}
\rho = (x|h) =4 \mbox{Re} \left [ \int^{f_{\rm max}}_{f_{\rm min}} \frac{\tilde{x} (f)  \tilde{h}^* (f)}{S_n(f)}  df \right ],
\label{snr}
\end{equation}
where $\tilde{x} (f)$ is the Fourier transform of the observed data, $\tilde{h} (f)$ is the template in the frequency domain, and $S_n(f)$ is the noise power spectrum density of a detector.
We set $f_{\rm max}=2048~{\rm Hz}$.
For $f_{\rm min}$, we set $f_{\rm min}=40~{\rm Hz} $ ($20~{\rm Hz}$) when the maximum $\Delta t_{\rm {echo}}$ of the search region is less (greater) than $0.4$~seconds.
This is because heavier detector-frame mass events, which have longer $\Delta t_{\rm {echo}}$, have lower merger-ringdown frequency.
The template is normalized so that $(h|h) =1$. 
We use the Welch's method to estimate the detector's noise power spectrum density \cite{welch,allen}.

By rewriting the frequency domain template as $\tilde{h}(f) = \tilde{h}(f; t_0 = 0) e^{-2 \pi i f t_0}$, we obtain $\rho$ in Eq.~\eqref{snr} as a function of $t_0$.
Here $t_0 $ is the initial time of the time domain template. 
We search for the maximum value of network SNR, denoted by $\rho_{\rm{event}}$, at $ t_0 =  t_{\rm merger} + \Delta t_{\rm echo}$ with an interval $\pm 0.01\, \overline{\Delta t}_{\rm echo} $, 
where $\overline{\Delta t}_{\rm echo}$ is the average value of $\Delta t_{\rm echo}$ in the search region determined for each event \cite{abedi}.

We use 90\% credible region of remnant mass and spin for the search range of each event.
The exact values are summarized in Appendix \ref{search}.
We use KAGRA Algorithmic Library ({\sc KAGALI}) to perform the analysis \cite{oohara}.

\subsection{Background estimation method}
The matched filter SNR becomes large when the template matches the signals in the data.
However, we can also obtain high SNR even when the template is  matched to short transient noise.
Therefore, since $\rho_{\rm event}$ itself cannot be evidence for the presence of signals, we evaluate the frequency of $\rho_{\rm event}$ as p-values comparing it with the background data.

Previously, we have used 4096 seconds data for each event to obtain p-values \cite{uchikata}.
However, 4096 seconds data are not available for some O3 events and the computational cost is high when using 4096 seconds data.
In this work, one of the purposes is to investigate whether the tendency of giving small p-values in O1 and O2 is the same for O3 events when the method in Ref.~\cite{abedi} is used. 
Therefore, we employ the similar method used in Ref.~\cite{abedi}. 
We use the later part of the data that are used for the signal search and select 1000 background segments with the interval $0.02\, \overline{\Delta t}_{\rm echo}$ from $8\, \overline{\Delta t}_{\rm echo}$ after the first echo (BG1).

In the above case, the data used for the background estimation are only $8\, \overline{\Delta t}_{\rm echo}$ apart from the first possible echo.
Furthermore, each background segment with the interval $0.02\, \overline{\Delta t}_{\rm echo}$ is adjoined.
If $0.02\, \overline{\Delta t}_{\rm echo}$ is too small, we may obtain multiple maximum SNRs from the same origin.
As a result, the effective number of the background segment can be less than 1000.
Therefore, we also consider the improved background estimation method (BG2).
We use the off-source data, which is described in the next subsection, and divide the background data into 25~ms segments and select the maximum SNR from the first $0.02\, \overline{\Delta t}_{\rm echo}$ interval in each 25~ms segment.

In both cases, we compute the maximum SNR in each segment by varying search parameters and count the number of background segments whose SNR exceeds SNR$_{\rm event}$.
 Then, p-values are given as the ratio of the above number to the total background segments as $p = \#(\rho \geq \rho_{\rm{event}} )/1000$.

\subsection{Data and events}

We select 34 black hole binary events with the false alarm rate smaller than $10^{-3} / {\rm yr}$ observed in O3\footnote{The updated false alarm rates of GW190421 and GW190521 exceed this threshold for the GstLAL search pipeline \cite{gwtc2-1}. However, inclusion of these events do not affect very much to overall p-value distributions, therefore, we also analyze these events in our analysis.  } , which is the same criteria used in Refs.~\cite{TGR-gwtc2,tgr-gwtc3}.
Furthermore, we select events that were observed by both LIGO Livingston and LIGO Hanford.
The data are obtained from the Gravitational Wave Open Science Center \cite{gwosc}.
We basically omit the last 6 digits of the event name except for GW190521\_074359, GW190828\_063405, and GW190828\_065509, where we rename them as GW190521\_07, GW190828\_063, and GW190828\_065, respectively.

We use the data from about 7 seconds before the binary merger time for all events and apply the Tukey window with a parameter $\alpha=1/8$ to the timeseries data.
We call this the on-source data.
We further use the data subsequent to the on-source data for the improved background estimation method, which we refer to as the off-source data\footnote{For GW191109, we skip the subsequent 64 seconds data since the subsequent data were noisy.}.
The duration of the data is 32 (64) seconds when the maximum $\Delta t_{\rm {echo}}$ of the search region is less (greater) than $0.4$~seconds (see Table \ref{search1} in Appendix \ref{search} for the search regions).
The sampling frequency of the data is 4096~Hz.

\section{Results}
\subsection{P-values for each event}

\begin{table*}[htb]
\begin{center}
\caption{P-values for each event when using the Simple model. P-values are provided from two different background estimation methods.}
\begin{tabular}{l l  l  l l  l  l l  l l  l l l  l l }
\hline
Event 	     	 & &  p-value  &         &Event 	 &	&  p-value &   		&Event 	&	& p-value & &  Event && p-value  \\  
			&  &BG1 / BG2 &      	& &	&BG1 / BG2 & 		                  &                 &  &BG1 / BG2  &	&	& &BG1 / BG2 \\ 
\hline
GW190408    	 &  &0.862 / 0.649     &   & GW190602    &    & 0.569 / 0.437 &     &  GW190915 &	& 0.685 / 0.599	 && GW200202	& &0.076 / 0.065 \\
GW190412    	 &  &0.561 / 0.414      &   & GW190706  &	 & 0.230 / 0.168  &    & GW190924 	 &      &0.915 / 0.830	& &GW200208	& &0.191 / 0.036	 \\
GW190421    	 &  &0.676 / 0.423	  & &  GW190707	& & 0.788 / 0.794   &  &GW191109   &      & 0.981 / 0.574	& &GW200219	& &0.880 / 0.734\\
GW190503    	 &  &0.132 / 0.090	  & &  GW190720   & & 0.073 / 0.074 &     &GW191129 	 &     & 0.161 / 0.198  &	& GW200224  & & 0.879 / 0.771\\
GW190512    	 & &0.219 / 0.691	   & &   GW190727 &	&0.955 / 0.844	 &      & GW191204	 &     & 0.992 / 0.963  &	& GW200225 & &0.777 / 0.551\\
GW190517    	 & & 0.299 / 0.119	  &   &  GW190728 	&    &0.661 / 0.472	&  &  GW191215  &    & 0.345 / 0.206 &	 &  GW200311 & &0.027 / 0.022\\
GW190519    	 &  &0.976 / 0.908  	 &  &  GW190814  &	& 0.939 / 0.732	     &  & GW191222 &	& 0.852 / 0.757 &	& GW200316 & & 0.652 / 0.399 \\ 
GW190521    	 & &0.101 / 0.126	  &   &GW190828\_063 & & 0.030 / 0.038   &  &GW200115  && 1.000 / 0.816	& & & &\\
GW190521\_07  & &0.825 / 0.699	  &  & GW190828\_065 & & 0.137 / 0.109	&  & GW200129 &  &0.086 / 0.091  & && &\\ 
 \hline
 \end{tabular}

\label{pvalue1}
\end{center}
\end{table*}

\begin{table*}[htb]
\begin{center}
\caption{P-values for each event when using the BHP model. The remnant spins of GW190814 and GW200115 are below the spin range used for the fit of the reflection rate. 
P-values are provided from two different backgound estimation methods.
}
\begin{tabular}{l l  l  l l  l  l l  l l  l l l  l l }
\hline
Event 	   &  	 &  p-value  &     &Event 	&	   &  p-value &                      &Event 		  & & p-value &         &   Event     & & p-value  \\ 
		 &	& BG1 / BG2 &	  &		&	   &BG1 / BG2 &                   &			  &  &BG1 / BG2       & 	&	     && BG1 / BG2 \\  \hline
GW190408 &     & 0.355 / 0.283  &    & GW190602  &  & 0.044 / 0.022    &   &  GW190915  & & 0.201 / 0.116	&   & GW200202	&&0.051 / 0.031 \\
GW190412   & 	 &  0.837 / 0.725  &    & GW190706 &   & 0.803 / 0.592	&  &GW190924   & &0.241 / 0.373	&   &GW200208	&& 0.299 / 0.059	 \\
GW190421    &	 &  0.488 / 0.280  &   &  GW190707&	  & 0.341 / 0.407	&  & GW191109   &  & 0.226 / 0.485  &   & GW200219	& &0.912 / 0.740\\
GW190503    &	 &  0.866 / 0.640  &   &  GW190720 &  & 0.449 / 0.459  &  &GW191129 	  & & 0.556 / 0.886   &     & GW200224  &  &0.952 / 0.740\\
GW190512    &	 &  0.192 / 0.502  &   &   GW190727 &  &0.897 / 0.744  &     & GW191204   &  & 0.972 / 0.355 	&   &   GW200225 & &0.132 / 0.052\\
GW190517    &	 &  0.129 / 0.066 &   &  GW190728 &	  &0.881 / 0.893   &     &GW191215    &  & 0.412 / 0.240	&    &GW200311 & &0.516 / 0.315\\
GW190519    &	 &  0.199 / 0.127  &   &    GW190814 &   & 0.994 / 0.781&     &GW191222   & & 0.378 / 0.574 	&   &GW200316 & &1.000 / 0.937 \\ 
GW190521    &	 &  0.886 / 0.356  &   &    GW190828\_063 & & 0.463 / 0.304  & &GW200115 	& & 0.814 / 0.517	 & & &&\\
GW190521\_07 & & 0.607	 / 0.196 &   & GW190828\_065 & & 0.499 / 0.341&	& GW200129   & &0.655 / 0.031  &&&&\\ 

\hline
 \end{tabular}

\label{pvalue2}
\end{center}
\end{table*}

We summarize p-values from both background estimation methods for each event in Tables \ref{pvalue1} and \ref{pvalue2} for the Simple and BHP models, respectively.
As mentioned in Sec.~II.~B, the reflection rate used in the BHP model is valid for $0.6 < \chi < 0.8$.
Therefore, the model is not valid for GW190814 and GW200115, since their remnant spins are below this range.
These events are possible and confident neutron star-black hole merger events, respectively \cite{gw190814,nsbh}.
P-values for these events are just shown as a reference in TABLE II.

In Ref.~\cite{echoGW190521}, the evidence of echo signals modeled by stimulated Hawking radiation for GW190521 was reported and it was mentioned that GW190521 shows exceptional significance among other various events \cite{Abedi:2022bph}.
GW190521 was the second most massive binary black hole merger event observed in O3, whose total mass exceeds $100$~$M_{\odot}$ \cite{gw190521,Abbott:2020mjq}.
In our study, p-values for the event given by the BG1 (BG2) method are 0.101 (0.126) for the Simple model and 0.886 (0.356) for the BHP model.
Since the echo models in our study depend on the merger-ringdown signal, it is natural to assume that the SNR of echo signals is proportional to the SNR of each binary merger event.
From this point, it is  reasonable that our study does not show significance of GW190521 compared to other events.
As shown in Appendix \ref{injection}, high significance of echo signals should appear for large SNR events such as GW190521\_07 and GW200129 rather than GW190521, if sufficiently large echo signals modeled in our study are present in the data.
In addition, the searched frequency range of echo signals reported in Refs.~\cite{echoGW190521,Abedi:2022bph} was lower than the estimated ringdown signal of the event, which is different from this study.

The smallest p-values obtained by the Simple model is 0.022 for GW200311 using the BG2 method.
This event has $\rho_{\rm event} = 4.9$.
The same p-value is obtained as the smallest by the BHP model using the BG2 method, but for GW190602 with $\rho_{\rm event} = 5.5$.
Since the total number of the events analyzed in this study is 34, it is reasonable that one of the events has p-value smaller than 0.03, even if the data are consistent with noise.

\subsection{Distribution of p-values}

\begin{figure*}[htb]
\includegraphics[scale=0.55]{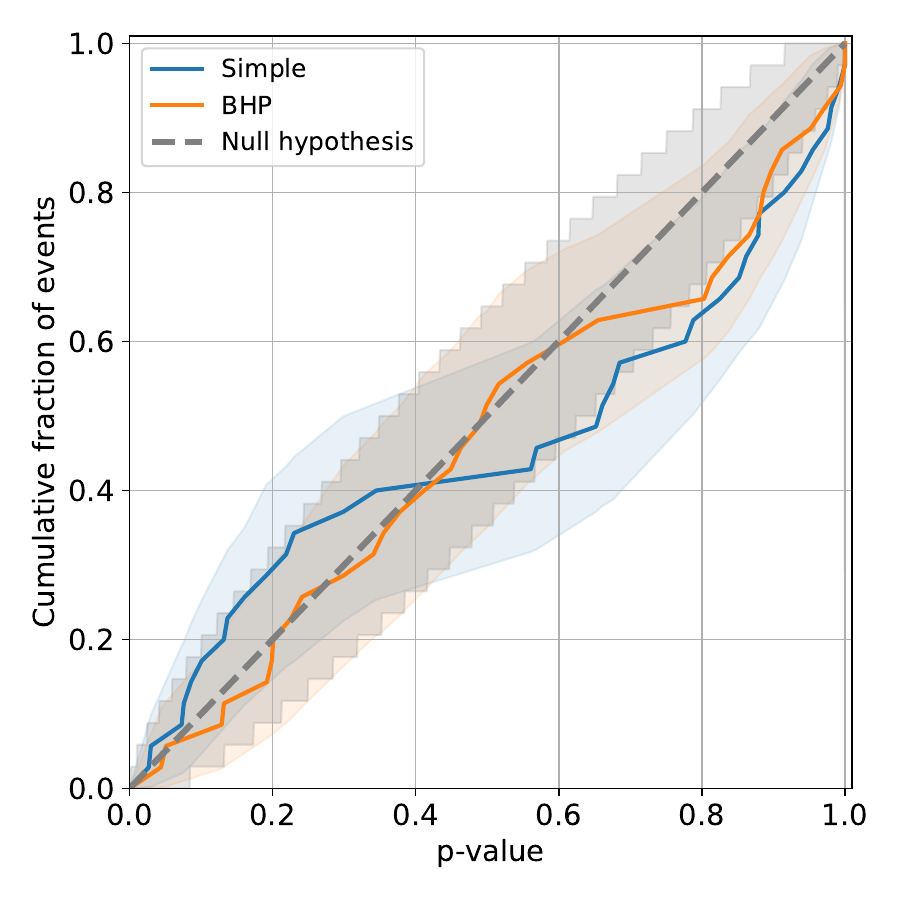}
\includegraphics[scale=0.55]{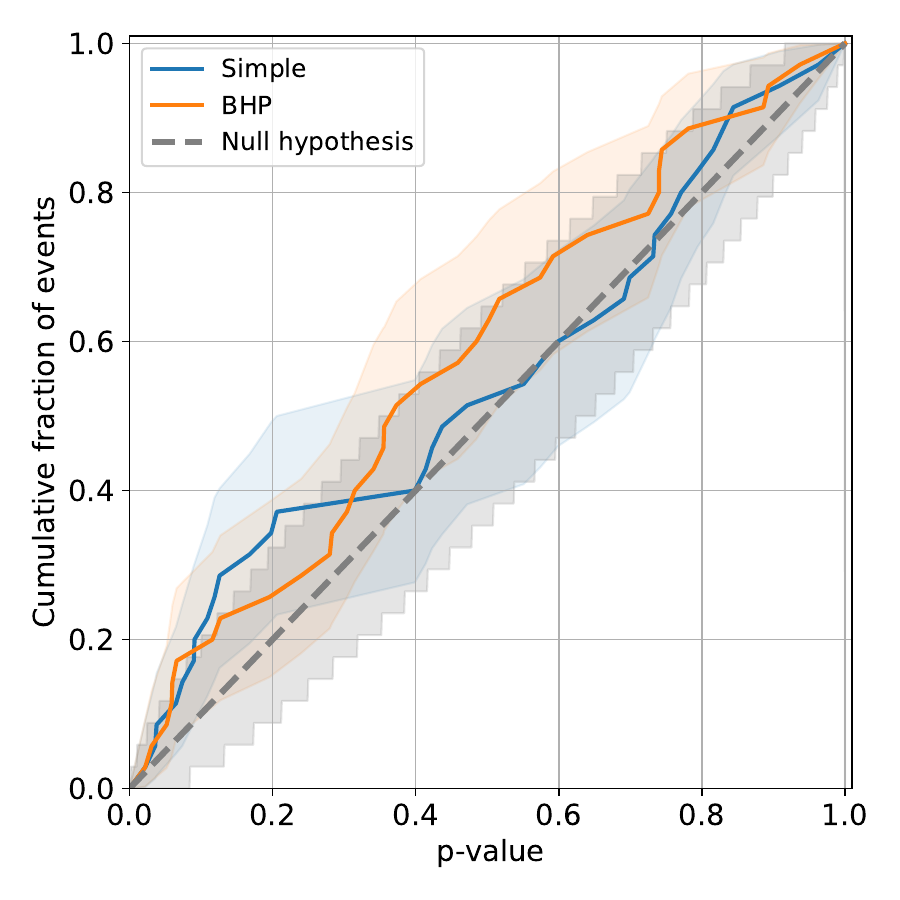}
\caption{Left: Distributions of a cumulative probability of p-values for the O3 events for the Simple model (blue) and the BHP model (orange). 
P-values are calculated from the first background estimation method (BG1).
Black dashed line corresponds to the uniform distribution (null hypothesis).
Shaded regions correspond to the 90\% error regions.
Right: The same as the left panel, but for the improved background estimation method case (BG2). }
\label{pp}
\end{figure*}

\begin{table*}[htb]
\begin{center}
\caption{List of mean p-values for total and classified cases.
SNR$_{\rm PI}$ is SNR of the post-inspiral part of each binary merger event used in Refs.~\cite{TGR-gwtc2,tgr-gwtc3}.}

\begin{tabular}{l   l  l    l l  l  l l l l  }
\hline
			   	&			& BG1 & 	&&	  				& BG2 & 			&		& \\
   Classification    	&No. of events	& Simple & 	  & BHP & 		& Simple & 		& BHP&  \\
   				&			& Mean & KS test & Mean & KS test &  Mean & KS test & Mean & KS test \\
   \hline
All				& 34	     		&0.545   &0.155 & 0.536&   0.344 & 0.453 &	0.214	& 0.418 &	0.318	 \\
SNR$_{\rm PI} > 10$ &21			&0.587   &	 0.122	 & 0.550 & 0.425&0.464 & 0.559 	&0.376 &   0.090    \\
$\Delta t_{\rm echo} \geq 0.4 $ s & 16 & 0.579&0.184	& 0.552 & 0.777&0.458 &  0.440	&0.330 &0.046 \\
$\Delta t_{\rm echo} < 0.4 $ s & 18  & 0.515& 	0.599	&0.550 & 	0.625 &0.449 & 0.525	& 0.495 & 0.602  \\
 \hline
 \end{tabular}
\label{meanP}
\end{center}
\end{table*}

When we have multiple events and simultaneously evaluate the significance, looking at p-value of each event is not an appropriate method.
Since the number of false positive events can increase if the number of events increases, i.e.~we can get many small p-value events.
In this study, we plot the distributions of a cumulative probability of p-value for both templates and compare with a uniform distribution in Fig.~\ref{pp}.
The left and right panels correspond to the results obtained by the BG1 and the BG2 methods, respectively.

If the data do not contain echo signals in all events, then we expect that p-values distribute uniformly.
The 90\% error regions are obtained by the Beta distribution for measured p-values and the binomial distribution is used for the error of the uniform distribution \cite{TGR-gwtc2,tgr-gwtc3}.
The error for the measured p-value distribution represents the uncertainty of true p-value from finite number of background segments. 
On the other hand, the error for the uniform distribution shows the uncertainty due to limited number of events.

Figure~\ref{pp} shows that the p-value distributions for the both models are consistent with the uniform distribution within 90\% error, irrespective of the background estimation method.
In Table~\ref{meanP}, we list mean p-values and p-values from the Kolmogrorov-Smirnov (KS) test for all events and classified events by SNR$_{\rm PI}$ or the length of $\Delta t_{\rm echo} $.
Here, SNR$_{\rm PI}$ is SNR of the post-inspiral part of each binary merger event used in Refs.~\cite{TGR-gwtc2,tgr-gwtc3}.
The KS test shows whether the distributions of p-values are consistent with the null hypothesis.
Small (large) KS test p-values mean that we can (cannot) reject the null hypothesis.
Typically, a threshold of 0.05 is used.
We can see that for most cases, p-values from the KS test is larger than 0.1 except for events with SNR$_{\rm PI} > 10$ or $\Delta t_{\rm echo} \geq 0.4 $~s when the BHP model and the BG2 method are used.
Mean p-values for these cases are less than 0.5, that is, the distributions slightly shift to the small p-values.
However, it should be noted that the number of events becomes small when a classification is applied.
We confirm that the distributions are consistent with the null hypothesis within 90\% errors for all cases, therefore, we cannot claim a deviation from the null hypothesis base on these results.
The results do not change very much even when we exclude higher false alarm rate events GW190421 and GW190521 nor include O1 and O2 events.

\section{Conclusions and discussions}
We have searched for gravitational wave echo signals from compact binary coalescences observed during O3.
We have performed the template-based search with two different models for echo signal templates: the Simple model and the BHP model.
The former is a repetition of the same waveform with a constant reflection rate at the surface, while the latter includes black hole perturbation effects in the reflection rate.
We have also applied two background estimation methods to compute p-values.
One is similar to the method used in Ref.~\cite{abedi}.
We have used this method to investigate whether it tends to give low p-values for O1 and O2 events.
Another method is similar to general methods used in the search of compact binary coalescences.
We compute p-value for all events and examine the distribution of p-values of all events.

We have not found any significant echo signals for both models for O3 events. 
All events show p-values larger than 0.02.
In Refs.~\cite{echoGW190521,Abedi:2022bph}, the exception of the significance of GW190521 has been reported, however, we have not found such exception of the event within our framework.
Although a deviation from the uniform distribution is observed for O1 and O2 events for the Simple model in our previous study, the distributions of p-values become consistent with the null hypothesis for both models within 90\% error region for O3 events, irrespective of the background estimation methods.
We have also confirmed that the conclusion does not change when we add O1 and O2 events.

Although we have not found any significant echo signals, the limited number of events analyzed still gives large error regions.
In the first three months of the ongoing O4, the LIGO detectors have detected more than 35 significant candidates for compact binary mergers\footnote{https://gracedb.ligo.org/superevents/public/O4/}, and we expect the LIGO and Virgo detectors to detect much more events in the remainder of the run. 
Therefore, it is necessary to analyze O4 events to give tighter constraint, which is left as a future work.

\section*{Acknowledgments}
N.~U.~thanks Jahed Abedi and Alex Nielsen for fruitful discussions.
This material is based upon work supported by NSF's LIGO Laboratory which is a major facility fully funded by the National Science Foundation.
This research has made use of data or software obtained from the Gravitational Wave Open Science Center (gw-openscience.org), a service of LIGO Laboratory, the LIGO Scientific Collaboration, the Virgo Collaboration, and KAGRA. LIGO Laboratory and Advanced LIGO are funded by the United States National Science Foundation (NSF) as well as the Science and Technology Facilities Council (STFC) of the United Kingdom, the Max-Planck-Society (MPS), and the State of Niedersachsen/Germany for support of the construction of Advanced LIGO and construction and operation of the GEO~600 detector. Additional support for Advanced LIGO was provided by the Australian Research Council. Virgo is funded, through the European Gravitational Observatory (EGO), by the French Centre National de Recherche Scientifique (CNRS), the Italian Istituto Nazionale di Fisica Nucleare (INFN) and the Dutch Nikhef, with contributions by institutions from Belgium, Germany, Greece, Hungary, Ireland, Japan, Monaco, Poland, Portugal, Spain. The construction and operation of KAGRA are funded by Ministry of Education, Culture, Sports, Science and Technology (MEXT), and Japan Society for the Promotion of Science (JSPS), National Research Foundation (NRF) and Ministry of Science and ICT (MSIT) in Korea, Academia Sinica (AS) and the Ministry of Science and Technology (MoST) in Taiwan.
This research has made use of a computing server VELA provided by ICRR, the University of Tokyo. 
This work was supported by JSPS KAKENHI Grant Numbers JP15H02087, JP16K05347, JP16K05356, JP20K03928, JP21H01082, JP21K03548, JP21K03582, JP23H00110, JP23K03381, and JP23K03432.

\appendix

\section{Injection study}
\label{injection}

\begin{figure}[htb]
\includegraphics[scale=0.55]{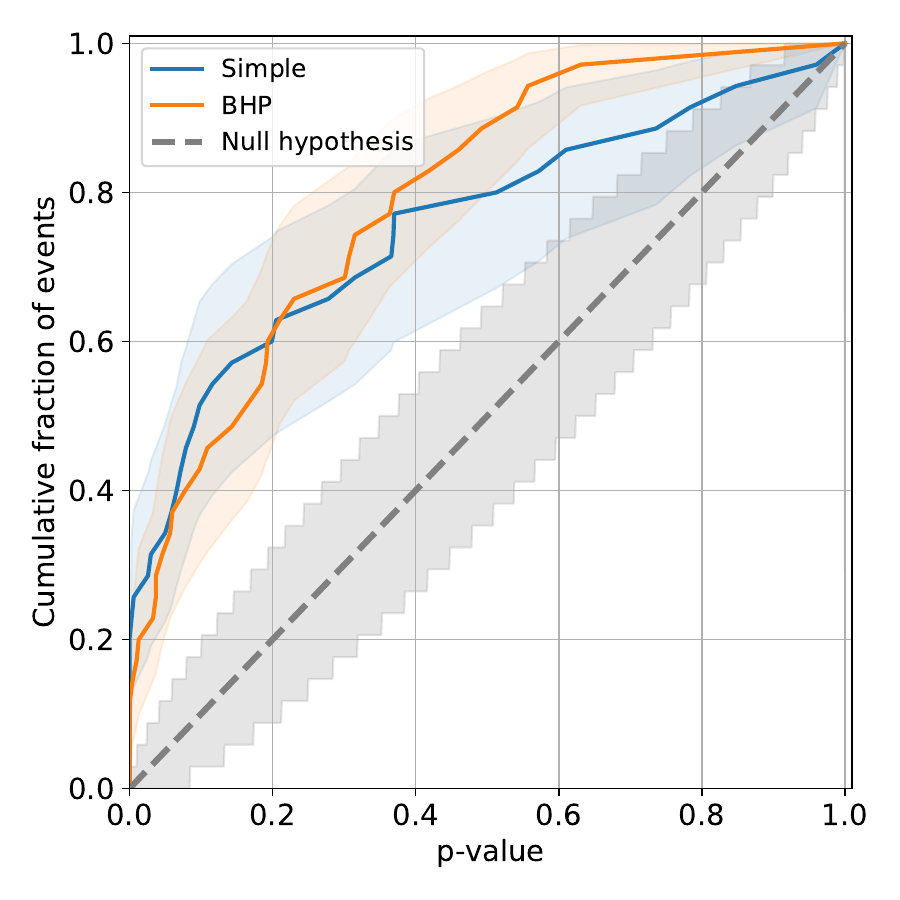}
\caption{Distributions of a cumulative probability of p-value for injected signals for the Simple model (blue) and the BHP model (orange). 
P-values are calculated from the improved background estimation method (BG2). }
\label{pp_inj}
\end{figure}

We apply our analysis method to simulated echo signals to validate the detectability of echo signals with sufficiently large SNR.
This is often called injection study.

We inject simulated echo signals into the off-source data that are not used for the background estimation.
We assume that the network SNR of echo signals is 30\% of the median network SNR of the binary mergers shown in Refs.~\cite{TGR-gwtc2,tgr-gwtc3} and SNR is the same for the both detectors.
We use the same background data used in Fig.~\ref{pp}.
We also set the start time of the first echo $t_{\rm echo}$ as $ t_{\rm echo} = t_{\rm merger} + \Delta t_{\rm echo} $.
The p-value distributions are shown in Fig.~\ref{pp_inj} for the improved background estimation method (BG2).
We see a clear deviation from the null hypothesis irrespective of echo models.
Mean p-values are 0.234 for the Simple model and 0.196 for the BHP model. 
The p-values from the KS test are $1.2 \times 10^{-6}$ and $6.6 \times 10^{-7}$ for the Simple and the BHP models, respectively.
We can confirm that these values are much smaller than those shown in Table~\ref{meanP}, which proves our method has sufficient ability to detect the injected signals.

\begin{table}[htb]
\begin{center}
\caption{A list of events with p-values less than $10^{-3}$ for the Simple model with injected and recovered values. The injected reflection rate is $\gamma = 0.8$ for all events. Network SNRs are shown for injected and recovered values. 
$\delta_t$ shows the relative error of the start time of the first echo.}
\begin{tabular}{l   l  l  l  l   l  r }
\hline 
Event          & SNR$^{\rm inj} $  &  $\Delta t_{\rm echo} ^{\rm inj}$ [s]    & SNR & $\Delta t_{\rm echo} $ [s] & $\gamma$  &$\delta_t$ [\%] \\
\hline
GW190412 & 5.7                   &  0.1831                                             &  7.8 &  0.1830                        & 0.90    & $-0.1$   \\
GW190519 & 4.7                    & 0.7582                                              & 6.6 &  0.7581                         & 0.90  &   $-0.01$\\
GW190521\_07 & 7.6              & 0.3959                                             & 7.9 &    0.3959                      &0.90 &         $0.03$\\
GW190814      &7.5                & 0.1048                                               &8.0  & 0.1048                       &  0.88 & $-0.1$   \\
GW190828\_063 & 4.7          & 0.3618                                                & 7.3 & 0.3619                       & 0.89 &   0.1 \\
GW191204   & 5.1               & 0.1002                                                  & 5.9 &   0.1003                   &0.80   &  $-0.3$   \\
GW200129   &8.0                 &0.3250                                               &  7.9 &  0.3249                      & 0.86 & $-0.8$\\
\hline
 \end{tabular}
\label{smallP}
\end{center}
\end{table}

\begin{table}[htb]
\begin{center}
\caption{A list of events with p-values less than $10^{-3}$ for the BHP model with injected and recovered values. 
Network SNRs are shown for injected and recovered values. 
$\delta_t$ shows the relative error of the start time of the first echo. }
\begin{tabular}{l   l  l  r  l  l  l  r r }
\hline 
Event  &  SNR$^{\rm inj} $  &  $a ^{\rm inj}$    &  $M^{\rm inj}  [M_{\odot} ]$ & &SNR & $a$ & $M [M_{\odot} ]$ &$\delta_t$ [\%]   \\
\hline 
GW190519 & 4.7              & 0.78                & 149.0                   &                   &6.6        &0.78 & 148.9     & $-0.03$       \\
GW190521\_07 & 7.7        &0.70                 & 87.3                      &              &  7.9       & 0.68  & 88.2 & 0.8 \\
GW190814 &7.5                   &0.30               &28.2                       &             &7.2          &0.28    &28.2 & $-1.0 $\\
GW200129 & 8.0         &      0.74                  &71.3                        &        &7.5            &0.72      &72.2  &  $-1.0$ \\

\hline
 \end{tabular}
\label{smallP2}
\end{center}
\end{table}

We also list the events with p-value less than $ 10^{-3}$ in Table~\ref{smallP} for the Simple model and Table~\ref{smallP2} for the BHP model.
Injected and recovered values are also shown in the tables.
These results show that in addition to detect signals, our method can recover injected values very well when the injected network SNR is larger than 7.
We also get similar results when the BG1 method is used.

\section{Search region}
\label{search}

\begin{table*}[htb]
\begin{center}
\caption{Durations of the data used for the search and background estimations, number of echoes $N_{\rm echo}$, and the search regions of $(a,M)$ with their corresponding $ \Delta t_{\rm echo}$ in our analysis for each event.}

\begin{tabular}{l   l  l l l       }
\hline
Event (duration)&  $N_{\rm echo}$  &$\chi$  &$M/ M_{\odot}$ &  $ \Delta t_{\rm echo} /s$ \\  \hline
GW190408  (32 s)    &  30 &$(0.58,0.73)$     & $(49.1,57.2)$ 	&$(0.2019,0.2555)$ \\
GW190412   (32 s)   & 30 & $(0.61,0.72)$     & $(38.2,47.4) $	&$(0.1564,0.2097) $\\
GW190421 (64 s)    &  20  &  $(0.57,0.81)$   & $(94.6,124)$ 	&$(0.3833,0.6125) $\\
GW190503  (64 s)   &  30 & $(0.53,0.78) $    & $(76.7,101) $	&$(0.3050,0.4767)  $\\
GW190512  (32 s)    &30 & $ (0.55,0.72) $     & $(40.7,50.2) $   &$(0.1623,0.2220)  $\\
GW190517 (64 s)	&  20&  $ (0.80,0.92) $   & $(74.5,89.6)$	 &$(0.3616,0.5789)  $\\
GW190519 (64 s) 	&  20 &  $(0.68,0.88) $   & $(133,165) $	&$(0.5761,0.9403)  $\\ 
GW190521 (64 s)	&   10& $(0.60,0.81) $    & $(235,281) $	&$(0.9766,1.4027) $\\
GW190521\_07  (64 s)& 30 & $ (0.64,0.75) $  & $(83.0,91.6)$ 	&$(0.3553,0.4365) $\\ 
GW190602  (64 s)	&  20& $ (0.60,0.82) $     & $(148,186) $	&$(0.6099,0.9367) $\\ 
GW190706  (64 s)	& 15& $(0.62,0.88) $       & $(150,195) $	&$(0.6263,1.1114) $ \\
 GW190707 (32 s)	& 30 & $ (0.61,0.68) $     & $(21.6,24.2) $	&$(0.0877,0.1028)$ \\
 GW190720  (32 s)	&  30 &$ (0.71,0.76) $     & $(22.5,26.6) $	&$(0.0949,0.1210) $\\
 GW190727 (64 s)	&  20 & $(0.63,0.85) $     & $(91.1,115) $	&$(0.3808,0.6108) $ \\
 GW190728   (32 s)  &  30  & $(0.67,0.76) $     & $(22.0,29.7) $	&$(0.0929,0.1355) $ \\
 GW190814   (32 s) &   30 & $(0.20,0.39) $     & $(23.5,32.8) $		&$(0.0857,0.1238) $ \\
 GW190828\_063 (64 s)& 30 & $(0.69,0.82)$   & $(71.3,82.9) $	&$(0.3091,0.4146) $ \\ 
 GW190828\_065  (32 s) & 30 & $(0.57,0.73)$  & $(38.6,49.5) $	&$(0.1548,0.2208) $ \\
 GW190915 (64 s)   &  30 & $(0.59,0.80)$ 	& $(69.5,83.0) $	&$(0.2835,0.4028)$ \\
 GW190924 (32 s)  & 30  & $(0.62,0.69) $	& $(14.1,16.8)$ 	&$(0.0574, 0.0712)$ \\
GW191109 (64 s)    &   30 & $(0.42,0.79)$ 	& $(120,154) $		&$(0.4630,0.7428) $\\
GW191129  (32 s)    & 30  &  $(0.64,0.72) $	& $(18.5,22.3) $	&$(0.0765,0.0975) $ \\ 
GW191204  (32 s)    &  30 &  $(0.7,0.76) $	& $(21.5,23.7) $	&$(0.0928,0.1077) $ \\ 
GW191215  (32 s)    & 30  & $(0.61,0.75) $	& $(52.5,60.6) $	&$(0.2157,0.2765) $\\
GW191222 (64 s)    & 30  &  $(0.56,0.75)$ 	& $(102,128)$ 		&$(0.4120,0.5884) $ \\
GW200115  (32 s)   &  30 & $(0.37,0.51) $	& $(5.9,9.6)$ 		&$(0.0218,0.0371) $ \\
GW200129 (32 s)    &  20 & $(0.68,0.79) $	& $(67.5,75.1)$ 	&$(0.2904,0.3597) $\\
GW200202  (32 s)   &  30  & $(0.65,0.72)$ 	& $(17.8,20.2)$ 	&$(0.0739,0.0884) $ \\
 GW200208   (64 s) &   30 & $(0.53,0.72)$ 	& $(78.4,97.8)$ 	&$(0.3119,0.4486 )$ \\
GW200219 (64 s)   & 30   & $(0.53,0.76)$ 	& $(87.0,113)$ 		&$(0.3465,0.5242) $ \\ 
GW200224  (64 s)  & 30   & $(0.66,0.8) $	& $(83.8,97.7)$ 	&$(0.3565,0.4755) $ \\
GW200225 (32 s)   &  30  & $(0.53,0.74) $	& $(35.8,42.3)$ 	&$(0.1412,0.1904)$ \\
GW200311 (32 s)  & 30   & $(0.61,0.76)$ 	& $(67.3,78.0) $	&$(0.2773, 0.3607)$ \\
GW200316 (32 s)  & 30  & $(0.66,0.74) $	& $(23.2,33.3) $	&$(0.0973, 0.1495)$ \\

 \hline
 \end{tabular}
\label{search1}
\end{center}
\end{table*}

In this appendix, we summarize the search regions for the durations of the data used for the search and background estimations, the number of echoes $N_{\rm echo}$, and $(a,M)$ of the remnant object with their corresponding $ \Delta t_{\rm echo}$ in Table~\ref{search1}.


\bibliography{echo_GWTC2_v5}
%

%

\end{document}